\documentstyle[preprint,tighten,aps,prb]{revtex}
\begin{document}
\preprint{cond-mat/9310067}
\draft
\title{Universality of weak localization in disordered wires}
\author{C. W. J. Beenakker}
\address{Instituut-Lorentz, University of Leiden,\\
P.O. Box 9506, 2300 RA Leiden, The Netherlands}
\date{August 1993}
\maketitle
\begin{abstract}
We compute the quantum correction $\delta A$ due to weak localization
for transport properties $A=\sum_{n}a(T_{n})$ of disordered
quasi-one-dimensional conductors, by integrating the
Dorokhov-Mello-Pereyra-Kumar equation for the distribution of the
transmission eigenvalues $T_{n}$. The result $\delta
A=(1-2/\beta)[\frac{1}{4}a(1)+ \int_{0}^{\infty}\!dx\,(4x^{2}+
\pi^{2})^{-1}a(\cosh^{-2}x)]$ is independent of sample length or mean
free path, and has a universal $1-2/\beta$ dependence on the symmetry
index $\beta\in\{1,2,4\}$ of the ensemble of scattering matrices. This
result generalizes the theory of weak localization for the conductance
to all linear statistics on the transmission eigenvalues.
\end{abstract}
%\newpage
\narrowtext
Weak localization is a quantum transport effect which manifests itself as a
magnetic-field dependent correction to the classical Drude conductance.
Discovered in 1979,\cite{And79,Gor79} it was the first-known quantum
interference effect on a transport property. (For reviews, see Ref.\
\onlinecite{reviews}.) At zero temperature, and in the quasi-one-dimensional
(quasi-1D) limit $L\gg W$ of a long and narrow wire (length $L$, width $W$),
the weak-localization correction to the conductance takes the universal
form\cite{Mel91,Note1}
\begin{eqnarray}
\delta G=\frac{2e^{2}}{h}\times\left\{\begin{array}{cl}
-\case{1}{3}&\hspace{1cm}(\beta=1)\\
0&\hspace{1cm}(\beta=2)\\
\case{1}{6}&\hspace{1cm}(\beta=4)
\end{array}\right.\label{deltaG1}
\end{eqnarray}
depending on the symmetry index $\beta$ of the ensemble of scattering matrices,
but independent of microscopic parameters as sample length $L$ or mean free
path $l$. If time-reversal symmetry is broken (e.g.\ by a sufficiently strong
magnetic field), then $\beta=2$ and $\delta G=0$. In the presence of both
time-reversal and spin-rotation symmetry, $\beta=1$ and $\delta G<0$. If only
the latter symmetry is broken (i.e.\ for strong spin-orbit scattering in zero
magnetic field), then $\beta=4$ and $\delta G>0$. The implication for the
magnetoconductance is that $G$ shows either a peak or a dip around zero field,
depending on the presence or absence of strong spin-orbit scattering. The peak
is precisely half as large as the dip.

The purpose of this paper is to demonstrate that the universality of the
weak-localization correction expressed by Eq.\ (\ref{deltaG1}) {\em is generic
for a whole class of transport properties}, of which the conductance is but a
special example. We consider a general transport property $A$ of the form
\begin{equation}
A=\sum_{n=1}^{N}a(T_{n}).\label{Adef}
\end{equation}
This is the definition of a {\em linear statistic\/} on the transmission
eigenvalues $T_{1},T_{2},\ldots T_{N}$. The word ``linear'' indicates that $A$
does not contain products of different $T_{n}$'s, but the function $a(T)$ may
well depend non-linearly on $T$. The conductance is a special case for which
$a(T)=(2e^{2}/h)T$ is linear in $T$ (Landauer's formula). Other examples of
linear statistics include the shot-noise power (with $a(T)$ a quadratic
function), the conductance of a normal-superconductor interface (with $a(T)$ a
rational function), and the supercurrent through a point-contact Josephson
junction (with $a(T)$ an algebraic function). In Ref.\ \onlinecite{Bee93} it
was shown that the theory of ``universal conductance fluctuations'' can be
generalized to all these linear statistics. Here we wish to establish such
generality for the theory of weak localization.\cite{Note2}

Our final result is a formula
\begin{equation}
\delta A=\left(1-\frac{2}{\beta}\right)\left(\frac{1}{4}a(1)+
\int_{0}^{\infty}\!\!dx\,\frac{a(\cosh^{-2}x)}{4x^{2}+
\pi^{2}}\right)\label{final}
\end{equation}
for the weak-localization correction $\delta A$ to the ensemble average
$\langle A\rangle=A_{0}+\delta A$ of an arbitrary linear statistic $A$ of the
form (\ref{Adef}). The term $\delta A$ is a quantum correction of order $N^{0}$
to the classical $\beta$-independent value $A_{0}$, which is of order $N$ (with
$N\gg 1$ being the number of scattering channels in the conductor). One easily
verifies that substitution of $a(T)=(2e^{2}/h)T$ into Eq.\ (\ref{final}) yields
the known result (\ref{deltaG1}), using
\[
\int_{0}^{\infty}\!\!dx\,(4x^{2}+\pi^{2})^{-1}(\cosh x)^{-2}=\frac{1}{12}.
\]
The fundamental significance of Eq.\ (\ref{final}) is that it demonstrates that
{\em all\/} linear statistics have a weak-localization correction which (i) is
independent of sample length or mean free path, and (ii) has a $1-2/\beta$
dependence on the symmetry index. In addition, Eq.\ (\ref{final}) reduces the
computation of the numerical value of the quantum correction to a quadrature,
regardless of the complexity of the function $a(T)$.\cite{Note3}

Starting point of the analysis is the Dorokhov-Mello-Pereyra-Kumar
equation\cite{Mel88}
\begin{equation}
\frac{\partial P}{\partial s}=
\frac{2}{\beta N+2-\beta}\sum_{i=1}^{N}
\frac{\partial}{\partial\lambda_{i}}\lambda_{i}(1+\lambda_{i})
J\frac{\partial}{\partial\lambda_{i}}J^{-1}P
\label{MPK}
\end{equation}
for the evolution of an ensemble of quasi-1D conductors of increasing length.
For each ratio $s=L/l$ the ensemble is characterized by the probability
distribution $P(\{\lambda_{n}\},s)$ of the set of variables
$\{\lambda_{n}\}=\lambda_{1},\lambda_{2},\ldots\lambda_{N}$. The
$\lambda$-variables are defined by $\lambda_{n}=(1-T_{n})/T_{n}$ in terms of
the transmission eigenvalues $T_{n}$. Since $T_{n}\in[0,1]$,
$\lambda_{n}\in[0,\infty)$. The $N$ degrees of freedom in Eq.\ (\ref{MPK}) are
coupled by the factor
$J(\{\lambda_{n}\})=\prod_{i<j}|\lambda_{i}-\lambda_{j}|^{\beta}$, which is the
Jacobian from the space of scattering matrices to the space of transmission
eigenvalues.\cite{Sto91} An exact solution of Eq.\ (\ref{MPK}) is
known,\cite{Bee93b} but only for the case $\beta=2$. This is of no use here,
since weak localization is absent for $\beta=2$. We therefore employ a
different method, which yields for any $\beta$ the eigenvalue density in the
large-$N$ limit. (This is the relevant limit for weak localization, which
requires $l\ll L\ll Nl$.) The key technical ingredient is an asymptotic
expansion published by Dyson more than twenty years ago,\cite{Dys72} but which
had remained largely unnoticed.

We seek to reduce Eq.\ (\ref{MPK}) to an equation for the density
$\rho(\lambda,s)=\langle\sum_{n}\delta(\lambda-\lambda_{n})\rangle$ of the
$\lambda$-variables. The brackets $\langle\cdots\rangle$ denote an average over
$\{\lambda_{n}\}$ with distribution $P(\{\lambda_{n}\},s)$. Multiplying both
sides of Eq.\ (\ref{MPK}) by $\sum_{n}\delta(\lambda-\lambda_{n})$ and
integrating over $\lambda_{1},\lambda_{2},\ldots\lambda_{N}$ one obtains an
equation
\begin{mathletters}
\label{rho1}
\begin{eqnarray}
&&\frac{\partial\rho}{\partial s}=\frac{2}{\beta N+2-\beta}\,
\frac{\partial}{\partial\lambda}\lambda(1+\lambda)
\left(\frac{\partial\rho}{\partial\lambda}-\beta I\right),\label{rho1a}\\
&&I(\lambda,s)=\int_{0}^{\infty}\!\!
d\lambda'\,\rho_{2}(\lambda,\lambda',s)(\lambda-\lambda')^{-1}
,\label{rho1b}
\end{eqnarray}
\end{mathletters}%
which contains an integral over the pair distribution function
\begin{equation}
\rho_{2}(\lambda,\lambda',s)=\langle\sum_{i\neq j}
\delta(\lambda-\lambda_{i})
\delta(\lambda'-\lambda_{j})\rangle.\label{rho2def}
\end{equation}
To close Eq.\ (\ref{rho1}) we use Dyson's asymptotic expansion\cite{Dys72}
\begin{equation}
\frac{I(\lambda,s)}{\rho(\lambda,s)}=\int_{0}^{\infty}\!\!
d\lambda'\,\frac{\rho(\lambda',s)}{\lambda-\lambda'}
+\frac{1}{2}\,\frac{\partial}{\partial\lambda}\ln\rho(\lambda,s)
+{\cal O}(N^{-1}).\label{Dyson}
\end{equation}
Substitution into Eq.\ (\ref{rho1a}) gives
\begin{eqnarray}
\frac{\partial\rho}{\partial s}=\frac{2}{\beta N+2-\beta}\,
\frac{\partial}{\partial\lambda}\lambda(1+\lambda)\rho
\frac{\partial}{\partial\lambda}\left(
\vphantom{\int_{0}^{\infty}}(1-\case{1}{2}\beta)\ln\rho
\right.\nonumber\\
\left.\mbox{}-\beta\int_{0}^{\infty}\!\!
d\lambda'\,\rho(\lambda',s)\ln|\lambda-\lambda'|\right).\label{rho2}
\end{eqnarray}

At this point it is convenient to switch to a new set of independent variables
$\{x_{n}\}$, defined by $\lambda_{n}=\sinh^{2}x_{n}$. Since
$T_{n}=(1+\lambda_{n})^{-1}$, one has $T_{n}=1/\cosh^{2}x_{n}$, with
$x_{n}\in[0,\infty)$. The ratio $L/x_{n}$ has the physical interpretation of a
channel-dependent localization length.\cite{Sto91} The density
$\tilde{\rho}(x,s)$ of the $x$-variables is related to $\rho(\lambda,s)$ by
$\tilde{\rho}=\rho\,d\lambda/dx=\rho\sinh 2x$. In terms of the new variables,
Eq.\ (\ref{rho2}) takes the form
\begin{eqnarray}
\frac{\partial\tilde{\rho}}{\partial s}=\frac{1}{2(N-\gamma)}\,
\frac{\partial}{\partial x}\tilde{\rho}
\frac{\partial}{\partial x}\left(\int_{0}^{\infty}\!\!
dx'\,\tilde{\rho}(x',s)u(x,x')
\right.\nonumber\\
\left.\vphantom{\int_{0}^{\infty}}\mbox{}-\case{1}{2}\gamma V(x)
-\case{1}{2}\gamma\ln\tilde{\rho}(x,s)\right),\label{rho3}
\end{eqnarray}
with the definitions $\gamma=1-2/\beta$, $V(x)=-\ln|\sinh 2x|$, $u(x,x')=-\ln
|\sinh^{2}x-\sinh^{2}x'|$. We need to solve Eq.\ (\ref{rho3}) to the same order
in $N$ as the expansion (\ref{Dyson}), i.e.\ neglecting terms of order
$N^{-1}$. To this end we decompose
$\tilde{\rho}=\tilde{\rho}_{0}+\delta\tilde{\rho}$, with $\tilde{\rho}_{0}$ of
order $N$ and $\delta\tilde{\rho}$ of order $N^{0}$. Substitution into Eq.\
(\ref{rho3}) yields to order $N$ an equation for $\tilde{\rho}_{0}$,
\begin{equation}
\frac{\partial\tilde{\rho}_{0}}{\partial s}=\frac{1}{2N}\,
\frac{\partial}{\partial x}\tilde{\rho}_{0}
\frac{\partial}{\partial x}\int_{0}^{\infty}\!\!
dx'\,\tilde{\rho}_{0}(x',s)u(x,x').\label{rho0eq}
\end{equation}
This is essentially the problem solved by Mello and Pichard,\cite{Mel89} who
showed that
\begin{equation}
\tilde{\rho}_{0}(x,s)=Ns^{-1}\,\theta(s-x),\label{rho0result}
\end{equation}
in the relevant regime $s\gg 1$, $s\gg x$. (The function $\theta(\xi)$ equals 1
for $\xi>0$ and 0 for $\xi<0$.) Eq.\ (\ref{rho0result}) implies that, to order
$N$, the $x$-variables have a uniform density of $Nl/L$, with a cutoff at $L/l$
such that $\int_{0}^{\infty}\!dx\,\tilde{\rho}_{0}=N$. In the cutoff region
$x\sim L/l$ the density deviates from uniformity, but this region is irrelevant
since the transmission eigenvalues are exponentially small for $x\gg 1$. One
can readily verify by substitution that the solution (\ref{rho0result})
satisfies Eq.\ (\ref{rho0eq}), using
\begin{equation}
\frac{\partial}{\partial x}\int_{0}^{\textstyle{s}}\!\!dx'\,u(x,x')=-2x,
\;{\rm for}\; s\gg 1,\: s\gg x.\label{integral}
\end{equation}

Now we are ready to compute the ${\cal O}(N^{0})$ correction
$\delta\tilde{\rho}$ to the density. Substituting
$\tilde{\rho}=\tilde{\rho}_{0}+\delta\tilde{\rho}$ into Eq.\ (\ref{rho3}), and
using Eqs.\ (\ref{rho0result}) and (\ref{integral}), we find
\begin{eqnarray}
\frac{\partial\delta\tilde{\rho}}{\partial s}=
\frac{1}{2s}\,\frac{\partial^{2}}{\partial x^{2}}\int_{0}^{\infty}\!\!
dx'\,\delta\tilde{\rho}(x',s)u(x,x')\nonumber\\
\mbox{}-\frac{1}{s}\frac{\partial}{\partial x}(x\delta\tilde{\rho})
-\frac{\gamma}{4s}\,\frac{\partial^{2}V}{\partial x^{2}}
-\frac{\gamma}{s^{2}}.\label{rho4}
\end{eqnarray}
The last term $\gamma/s^{2}$ on the r.h.s.\ is a factor $s$ smaller than the
other terms, and may be neglected for $s\gg 1$. Eq.\ (\ref{rho4}) thus has the
$s$-independent solution $\delta\tilde{\rho}(x)$ satisfying
\begin{eqnarray}
&&\frac{1}{2}\,\frac{d^{\,2}}{dx^{2}}\int_{0}^{\infty}\!\!
dx'\,\delta\tilde{\rho}(x')\ln|\sinh^{2}x-\sinh^{2}x'|\nonumber\\
&&\hspace{1cm}\mbox{}+\frac{d}{dx}[x\delta\tilde{\rho}(x)]
=\frac{\gamma}{4}\,\frac{d^{\,2}}{dx^{2}}\ln|\sinh 2x|.\label{rho5}
\end{eqnarray}

It remains to solve the integro-differential equation (\ref{rho5}). This can be
done analytically by means of the identity\cite{thanks}
\begin{eqnarray}
&&\int_{0}^{\infty}\!\!
dx'\,f(x')\ln|\sinh^{2}x-\sinh^{2}x'|=\nonumber\\
&&\hspace{2cm}\int_{-\infty}^{\infty}\!\!
dx'\,f(|x'|)\ln|\sinh(x-x')|,\label{identity}
\end{eqnarray}
which transforms the integration into a convolution. The Fourier transform then
satisfies an ordinary differential equation, which is easily solved. The result
is
\begin{equation}
\delta\tilde{\rho}(x)=(1-2/\beta)[\case{1}{2}\delta(x)+
(4x^{2}+\pi^{2})^{-1}],\label{deltarho}
\end{equation}
as one can also verify directly by substitution into Eq.\ (\ref{rho5}). The
correction (\ref{deltarho}) to the uniform density (\ref{rho0result}) takes the
form of a deficit (for $\beta=1$) or an excess (for $\beta=4$), concentrated in
the region $x\lesssim 1$. For $\beta=2$ there is no ${\cal O}(N^{0})$ deviation
from uniformity. The existence of a $\beta$-dependent density excess or deficit
in the metallic regime was anticipated by Stone, Mello, Muttalib, and
Pichard,\cite{Sto91} from the $\beta$-dependence of the localization length in
the insulating regime. However, as emphasized by these authors, their argument
is simply suggestive and needs to be made quantitative. Eq.\ (\ref{deltarho})
does that.

The weak-localization correction $\delta A$ follows upon integration,
\begin{equation}
\delta A=\int_{0}^{\infty}\!\!dx\,\delta\tilde{\rho}(x)\,a(1/\cosh^{2}x).
\label{deltaArho}
\end{equation}
Combination of Eqs.\ (\ref{deltarho}) and (\ref{deltaArho}) finally gives the
formula (\ref{final}) for the weak-localization correction to the ensemble
average of an arbitrary linear statistic, as advertized in the introduction.

We conclude with an illustrative application of Eq.\ (\ref{final}), to the
conductance $G_{\rm NS}$ of a disordered normal-metal--superconductor (NS)
junction. This transport property is a linear statistic for zero magnetic
field,\cite{Bee92}
\begin{equation}
G_{\rm NS}=\frac{2e^{2}}{h}\sum_{n=1}^{N}\frac{2T_{n}^{2}}{(2-T_{n})^{2}}.
\label{GNS}
\end{equation}
In a semi-classical treatment, the ensemble average $\langle G_{\rm NS}\rangle$
is just the Drude conductance --- unaffected by Andreev reflection at the NS
interface. However, the quantum correction $\delta G_{\rm NS}$ due to weak
localization is enhanced by Andreev reflection.\cite{Bee92} Previously, there
was no method to calculate $\delta G_{\rm NS}$.\cite{Note3} Now, using Eq.\
(\ref{final}) one computes (for $\beta=1$) $\delta G_{\rm
NS}=-(1-4\pi^{-2})(2e^{2}/h)$, which exceeds the result $\delta
G=-\frac{1}{3}(2e^{2}/h)$ in the normal state by almost a factor of two. The
experimental observation of the enhancement of weak localization by Andreev
reflection has recently been reported.\cite{Len93}

In summary, we have shown that the universality of the weak-localization effect
in disordered wires is generic for a whole class of transport properties, viz.\
the class of linear statistics on the transmission eigenvalues. A formula has
been derived which permits the computation of the weak-localization correction
in cases that previous methods were not effective. This quantum correction is
independent of sample length or mean free path, and has a $1-2/\beta$
dependence
on the symmetry index, for all linear statistics.

Discussions with B. Rejaei and F. L. J. Vos are gratefully acknowledged. This
research was supported in part by the Dutch Science Foundation NWO/FOM.

\end{document}